\documentclass[10pt]{article}

\usepackage{psfrag}
\usepackage{amsmath}
\usepackage{amssymb}
\usepackage{textcomp}		
\usepackage{array}		
\usepackage{multirow}		
\usepackage{rotating}		
\usepackage{graphicx}

\begin{document}

\title{Exact asymptotic expansions for thermodynamics of the hydrogen gas in
the Saha regime}

\author{A. Alastuey\thanks{Laboratoire de Physique, ENS Lyon, CNRS, 
46 all\'ee d'Italie, 69364 Lyon Cedex 07, \textsc{France}.}
\ and 
\ V. Ballenegger\thanks{Institut UTINAM, Universit\'e 
de Franche-Comt\'e, CNRS, 16 route de Gray, 25030 Besan\c{c}on 
Cedex, \textsc{France}.}}
\date{} 

\maketitle

\begin{abstract}
\begin{footnotesize}
We consider the hydrogen quantum plasma in the Saha regime,
where it almost reduces to a partially ionized atomic gas. We
briefly review the construction of systematic expansions of thermodynamical
functions beyond Saha theory, which describes an ideal mixture
of ionized protons, ionized electrons and hydrogen atoms in their
ground-state. Thanks to the existence of rigorous results, we first
identify the simultaneous low-temperature and low-density limit
in which Saha theory becomes asymptotically exact. Then, we argue that
the screened cluster representation is well suited for
calculating corrections, since that formalism
accounts for all screening and recombination phenomena at work in a
more tractable way than other many-body methods. We sketch the
corresponding diagrammatical analysis, which leads to an
exact asymptotic expansion for the equation of state.
That scaled low-temperature expansion improves the
analytical knowledge of the phase diagram. It also
provides reliable numerical values over a rather
wide range of temperatures and densities, as confirmed
by comparisons to quantum Monte Carlo data.
\end{footnotesize}
\end{abstract}

\section{Introduction}

Obtaining asymptotically exact formulae for the equation of state of quantum Coulomb matter is important, both at a theoretical level and for practical applications. They provide a better understanding of basic phenomena like molecular recombination and screening in the framework of statistical mechanics. Such formulae are free from any intermediate phenomenological modelization and uncontrolled approximation. They provide moreover reliable and accurate data in some range of thermodynamical parameters. Exact expansions are of particular interest for hydrogen described as a gas of quantum protons
and electrons interacting \textsl{via} the Coulomb potential. Indeed,
thanks to its relative simplicity, analytical calculations can
be carried out further than for heavier species. In practice, the corresponding
expansions are quite useful since hydrogen is the most abundant element in the universe.
Astrophysicists need accurate equations of state over a wide range of
temperatures and densities, including the so-called Saha regime where
hydrogen reduces to a partially ionized atomic gas.

\bigskip

In the $50$'s~\cite{GellMont}, a first kind of asymptotic expansion
was derived for an electron gas at high densities, which behaves as
a free Fermi gas in a first approximation. Corrections can be computed in a systematic way
in the framework of standard many-body perturbation theory~\cite{FettMar},
where the small-expansion parameter is the charge of the electrons.
A second kind of asymptotic expansions, namely the familiar
virial expansions, where constructed in the $60$'s~\cite{Ebe},
for two- or more component systems, including quantum hydrogen.
In such expansions, temperature $T$ is fixed at a non-zero value,
and numerical densities $\rho_{\alpha}$ for species $\alpha$
are driven to zero. For hydrogen, we set $\rho=\rho_p=\rho_e$.
At lowest order, the system
behaves as an ideal mixture of nuclei and electrons, as
rigorously proved in Ref.~\cite{Lebo}. Corrections are represented
by series involving integer and half-integer powers
of the $\rho_{\alpha}$'s, as well as logarithmic terms. Corresponding calculations have
been first performed up to order $\rho_{\alpha}^2$~\cite{Kraeft}, by using
the effective-potential method introduced
by Morita~\cite{Morita}. Further
corrections of order $\rho_{\alpha}^{5/2}$ have been derived in the
$90$'s within another formalism based on the path integral
representation~\cite{AlaPer}, and retrieved later by
Ebeling-Morita's method~\cite{Khalbaum}.

\bigskip

Above expansions are suited for regimes where the systems are
almost fully ionized. They cannot describe
the Saha regime for hydrogen, where a finite fraction of electrons and
protons recombine into atoms in their groundstate. Then,
according to the familiar Saha theory~\cite{Saha}, the system is expected
to behave as an ideal mixture of ionized protons,
ionized electrons and atoms $H$. The construction
of suitable asymptotic expansions requires first to
identify, if it exists, a regime of thermodynamical parameters
where Saha theory becomes asymptotically exact.
As described in Section 2, that regime is obtained
by setting $T \to 0$, while ratio $\rho/\rho^{\ast}$ is kept fixed
with temperature-dependent density $\rho^{\ast}$ given
by expression \eqref{rhostar}.

\bigskip

Once the proper limit which defines the Saha regime has been
identified, the construction of systematic expansions
beyond Saha theory requires a formalism which
accounts for all recombination and screening phenomena at work.
For that purpose, the screened cluster
diagrammatical representation~\cite{Alastuey03} is particularly
adequate, as briefly described in Section 3. The corresponding
analysis of all involved graphs, provides the so-called SLT
expansion of pressure $P$, \textsl{i.e.}
\begin{equation}
\label{SLTpressure}
\beta P/\rho^{\ast} = \beta P_{Saha}/\rho^{\ast} +
\sum_{k=1}^{\infty} b_k(\rho/\rho^{\ast}) \alpha_k(\beta)~,
\end{equation}
where Saha pressure  $P_{Saha}$ is given by formula
\eqref{Sahapressure} in section~2. Functions $b_k(\rho/\rho^{\ast})$ only depend on ratio
$\rho/\rho^{\ast}$, while temperature dependent functions
$\alpha_k(\beta)$ decay exponentially fast when $T$ vanishes,
$\alpha_k(\beta) \sim \exp(-\beta \delta_k)$ except for possible
multiplicative powers of $\beta$. Expansion \eqref{SLTpressure}
is ordered with respect to the decaying rates
$0 < \delta_1 < \delta_2 < ...$ of the $\alpha_k(\beta)$'s functions. The first five corrections
computed in Ref.~\cite{Alastuey08} are schematically presented in Section 4.
They account for non-ideal phenomena such as plasma polarization,
shift in the atomic energy levels, interactions
between ionized charges and atoms, and also formation of
molecules $H_2$ or ions $H^-$ and $H_2^+$. Further
corrections $k \geq 6$ decay exponentially faster than
$\exp(\beta E_H)$, where $E_H=-me^4/(2\hbar^2)$ is the atomic
groundstate energy and
$m = {m_p m_e}/(m_p + m_e)$ is the reduced mass
for the two-body electron-proton problem.
Along a given low-temperature isotherm, we also study
the behaviour of the $b_k(\rho/\rho^{\ast})$'s.
SLT expansion \eqref{SLTpressure} then appears
as a partial infinite resummation of ordinary virial
expansions at low densities $\rho \ll \rho^{\ast}$. Expansion \eqref{SLTpressure} remains valid at intermediate ($\rho \sim \rho^\ast$) and large ($\rho > \rho^\ast$) densities, but breaks down at too large densities $\rho \gg \rho^{\ast}$ because molecular recombination becomes then prominent.

\bigskip

\begin{figure}[h]
\begin{center}
\includegraphics[width=13cm,angle=0]{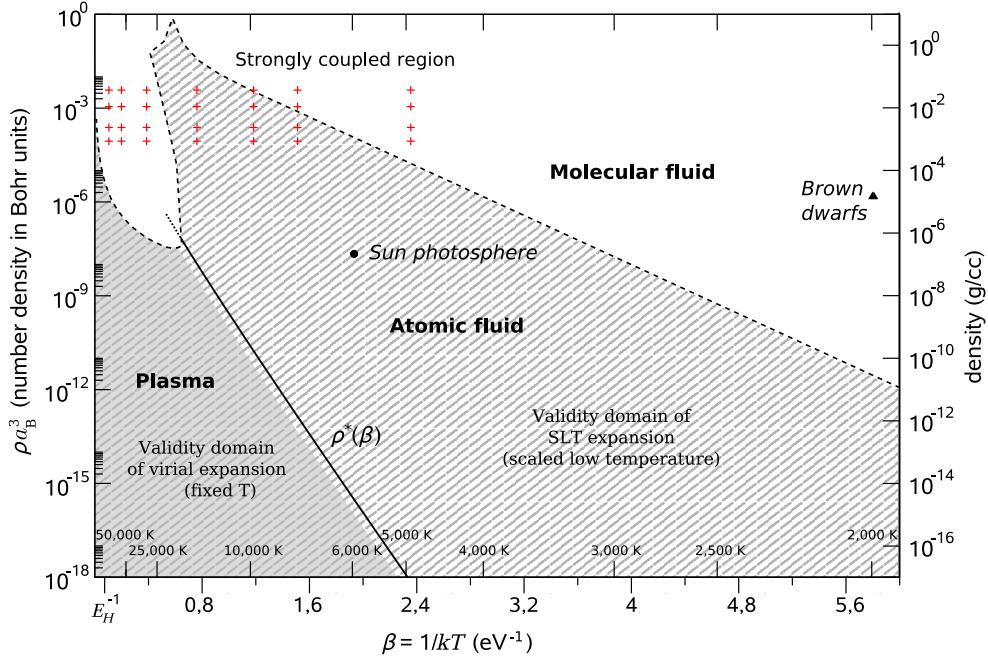}
\caption{Validity domain of truncated SLT equation of state\label{fig_validitydomain}}
\end{center}
\end{figure}

As usual for asymptotic series, keeping only the first few terms
of SLT expansion \eqref{SLTpressure} should lead to
an accurate equation of state, provided that
thermal energy $k_BT$ is smaller than Rydberg energy $|E_H|$. In
Section 5, we give a flavour of numerical calculations based
on the truncation of \eqref{SLTpressure} up to term $k=5$
included~\cite{AlaBalCor09}. Because of the relatively
large temperature scale $|E_H|/k_B \simeq 150 000K$, and
of the occurrence of exponentially decaying factors, that truncated
equation of state is reliable over a rather wide range of
thermodynamic parameters, as confirmed by
comparisons to quantum Monte Carlo simulations by Militzer and
Ceperley~\cite{Militzer01}. This allows us to
introduce a semi-empirical criterion for the convergence of
SLT expansions, which provides the validity domain of the
corrected EOS in the temperature-density plane. Our formulae
should be particularly useful in
physical situations where deviations to Saha theory
play an important role even if they remain small. For
instance, a very accurate EOS is needed for interpreting
recent seismology measurements in the Sun~\cite{Daeppen06}.
Our corrected EOS should be useful since,
according to usual models, the Sun adiabat lies in the previous validity
domain (see Fig.~\ref{fig_validitydomain}). However, notice
that an accurate description of that adiabat requires
to take into account heavier species like
helium, carbon, nitrogen and oxygen. If exact calculations
become much more complicated, a simple account of obvious ideal
contributions, in particular for less abundant species,
should be sufficient for significantly improving a
pure-hydrogen EOS.

\bigskip

We stress that our approach does not provide unambiguous
definitions of neither free and bound charges,
nor ionization rates. As commented in Section~\ref{Sct4},
this does not cause any trouble as far as thermodynamic quantities
are concerned. Other quantities like conductivity
or opacity cannot be obtained within the present formalism.
Usually, such quantities are computed within the
framework of the chemical picture, where atoms and molecules
are introduced phenomenologically as preformed entities.
A first-principles description is of course
possible in principle, but it becomes quite cumbersome since
it would require a rigorous
introduction of either real-time evolution or
coupling to radiation.

\section{The hydrogen gas in the Saha regime}	\label{Sct2}

\subsection{The physical picture}

Within the physical picture, a hydrogen gas is viewed as a system
of quantum point particles which are either protons or electrons,
interacting via the instantaneous Coulomb potential $v(r)=1/r$.
Protons and electrons have respective charges, masses, and spins,
$e_p=e$ and $e_e=-e$, $m_p$ and $m_e$, $\sigma_p=\sigma_e=1/2$.
In the present non-relativistic limit,
the corresponding Hamiltonian for $N=N_p+N_e$ particles reads
\begin{equation}
H_{N_p,N_e} = -\sum_{i=1}^N {\hbar^2 \over 2m_{\alpha_i}} \Delta_i +
{1 \over 2}\sum_{i \neq j} e_{\alpha_i} e_{\alpha_j} v(|\textbf{x}_i - \textbf{x}_j|)
\label{deuxun}
\end{equation}
where $\alpha_i=p,e$ is the species of the $i$th particle and
$\Delta_i$ is the Laplacian with respect to its position $\textbf{x}_i$.
The system is enclosed in a box with volume $\Lambda$,
in contact with a thermostat at temperature $T$ and a reservoir of particles
that fixes the chemical potentials equal to $\mu_p$ and $\mu_e$
for protons and electrons respectively. Because the infinite system maintains
local neutrality $\rho_p=\rho_e$ in any fluid phase, the bulk equilibrium
quantities depend in fact solely on the mean
\begin{equation}
\mu=(\mu_p + \mu_e)/2,
\label{deuxquatre}
\end{equation}
while the difference $\nu=(\mu_e-\mu_p)/2$ is not relevant as
rigorously proved in ref.~\cite{LieLeb}.

\subsection{Identification of the scaled limit}

In the so-called Saha regime, a finite fraction of protons and electrons
combine into hydrogen atoms, forming a partially ionized hydrogen gas.
That regime is attained when several conditions are met. The temperature must be sufficiently low
so that atoms can form, namely $kT \ll |E_H|$. The density must be sufficiently low as well
so that atoms maintain their individuality  thanks to $a \gg a_B$,
where $a$ is the mean inter-particle distance and $a_B$ is the Bohr radius.
If the density becomes too low, atoms dissociate by entropy, while if
it becomes too large, they recombine into molecules $H_2$.
According to those simple considerations, both $T$ and $\rho$
must go to zero, in a related way. The precise form of that
relation can be inferred from a rigorous analysis
in the grand-canonical ensemble devised by
Macris and Martin~\cite{Macris90}, who extended Fefferman's work
on the atomic phase of the hydrogen plasma~\cite{Fefferman85}.
They introduce a scaling limit where the temperature~$T$ goes to zero, while
the average chemical potential $\mu$ of protons and electrons
approaches the ground-state energy $E_H$ with a definite slope~\cite{Macris90}.
Then, they proved that pressure $P$, within that scaling limit,
tends to its Saha expression $P_{Saha}$, which
describes an ideal mixture of hydrogen atoms, ionized protons
and ionized electrons.

\bigskip

In terms of temperature and density, the previous scaling limit
can be rephrased as a low temperature expansion at fixed ratio 
$\rho/\rho^{\ast}$ [see eq. (1)], where $\rho^{\ast}$ is the temperature-dependent density
\begin{equation}
 \rho^* =\frac{\exp (\beta E_H)}
 {2 (2\pi \lambda_{pe}^2)^{3/2}}\quad
 \text{with} \quad
\lambda_{pe} = (\beta \hbar^2/m)^{1/2}~.
\label{rhostar}
\end{equation}
Notice that density vanishes exponentially fast when $T$ is sent to zero.
This ensures the proper energy-entropy balance which keeps
a finite ionization rate that is entirely determined by
the fixed ratio $\rho/\rho^{\ast}$. Pressure $P$ in units of
$\rho^{\ast}k_BT$ tends to Saha formula
\begin{equation}
\beta P_{Saha}/\rho^{\ast} = \rho/\rho^{\ast} +
(1+2\rho/\rho^{\ast})^{1/2} - 1
\label{Sahapressure}
\end{equation}
apart from exponentially small terms when $T \to 0$.
The respective behaviours of Saha pressure \eqref{Sahapressure},
$P_{Saha} \sim 2 \rho$ for $\rho \ll \rho^{\ast}$, and
$P_{Saha} \sim  \rho$ for $\rho \gg \rho^{\ast}$ clearly
illustrate that $\rho^{\ast}$ is a cross-over density
between full ionization and full recombination.

\section{Construction of SLT expansion}	\label{Sct3}

\subsection{Introduction of a suitable formalism}

Corrections to Saha pressure \eqref{Sahapressure} involve
interactions between ionized charges and atoms, as well as
formation of ions and molecules. Standard many-body theory is not
well-suited for taking into account recombination, since that
mechanism is not perturbative with respect to the charge. For instance,
an infinite number of Feynman ladder graphs must be resummed
for describing a single atom $H$.

The effective-potential method,
which amounts to introduce a classical equivalent
system of point particles with many-body effective interactions,
is \textsl{a priori} more efficient for dealing with recombination. Indeed,
$n$-body effective interactions are inferred from
$n$-body quantum Gibbs factors which do account for recombination
of $n$ particles at short distances. The contributions of
two-body effective interactions can be analyzed within
standard methods of classical statistical mechanics. In particular,
that feature has been exploited for computing virial expansions up to
order $\rho^2$~\cite{Ebe}, where atomic contributions appear.
Ionic and molecular recombination are embedded in three- and
four-body effective interactions. Unfortunately, the analysis of
the corresponding contributions becomes rather cumbersome,
in particular because no standard classical tool is available.

Above drawbacks of both standard many-body theory and effective-potential
method, clearly emphasize the need for a formalism more appropriate
to deal with Saha regime. The ACTEX method, introduced
by Rogers~\cite{Rogers81} in the 70's, is intended to account for the
formation of chemical species in the framework of the
physical picture. That approach starts from the usual activity
expansion of thermodynamical quantities in the grand-canonical
ensemble. Despite its rather successful predictions
at moderate densities and temperatures, that approach cannot
be applied here as it stands, because ACTEX
series are not exactly reorganized \textsl{via} a
systematic treatment of both recombination and screening.
In fact, quantum Gibbs factors involved in activity series do not factorize
as products of two-body counterparts, like in the case of
classical charges. Resummations,
which are crucial for taking into account screening effects, are very hard to handle in the usual quantum activity series.

The difficult task of controlling simultaneously recombination and screening effects in quantum activity series can be accomplished thanks to the Feynman-Kac path integral representation.
Within that formalism, the genuine quantum system of
point protons and electrons is shown to be
equivalent to a classical gas of extended loops~\cite{Magic}.
Thermodynamic quantities of hydrogen are then represented by
activity series in the world of loops, which can be suitably
rearranged as described below.

\subsection{The screened cluster representation}

Since loops are classical objects with two-body interactions,
standard Mayer diagrammatical methods can be applied. In particular,
the one-body loop density is represented by a series of
Mayer-like graphs in the grand-canonical ensemble. Usual points are replaced by loops,
loop fugacities are simply related to particle fugacity $z=\exp (\beta \mu)$,
while Mayer bonds are built with the loop-loop interaction potential.
Since that potential behaves as the Coulomb interaction at large distances,
Mayer graphs are plagued with long-range divergences. Such divergences
are systematically removed \textsl{via} chain resummations, which amount
to introduce a screened potential~\cite{BalMarAla}. Contrary to the familiar classical Debye potential which decays
exponentially fast, that quantum potential decays only as
$1/r^3$ at large distances $r$. However, at low densities,
it reduces to its classical Debye counterpart plus small
corrections. At the same time,
the whole series is reorganized in terms of particle clusters.
Eventually, particle density $\rho$, obtained by integrating
loop density over all possible shapes, is exactly rewritten as
the following diagrammatical series~\cite{Alastuey03}
\begin{equation}
\psfrag{F1}{\footnotesize$\Phi$}
\psfrag{F2}{\footnotesize$\Phi^2$}
\psfrag{F3}{\footnotesize$\Phi^3$}
\rho =  \raisebox{-3mm}{\includegraphics[scale=0.6]{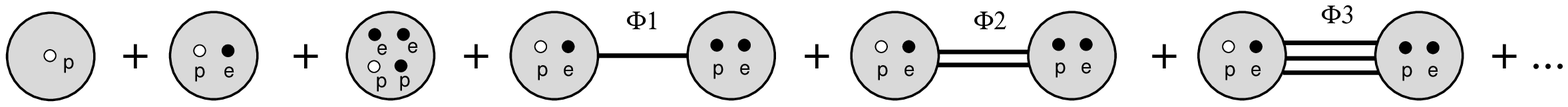}}
\label{SCEdensity}
\end{equation}
The corresponding graphs are constructed with topological rules close
to that of ordinary Mayer graphs, except for some exclusion constraints
avoiding double counting. Usual points
are now replaced by particle clusters. The statistical weight of a given
cluster involves particle fugacities, as well as screened
interactions. Two clusters
can be connected by a single bond, which is either $-\beta\Phi$,
$\beta^2\Phi^2/2$, or $-\beta^3\Phi^3/6$, where $\Phi$ is the
screened interaction between those clusters.


Expansion~\eqref{SCEdensity} accounts in a fully consistent way for all effects of interactions in the system at finite density and finite temperature. The various phenomena at work are embedded in well-defined graphs. For instance,
the first graphs shown in \eqref{SCEdensity} account respectively for a single ionized proton, formation of an atom $H$ and of a
molecule $H_2$, and interactions between two atoms. 
As the scaling limit defines quite diluted conditions,
only a few simple graphs in the screened cluster expansion are expected to contribute to the first corrections to Saha theory.

\subsection{Behaviour of graphs in the scaling limit}

The scaling limit defined in Section 2.2 within
the grand-canonical ensemble, can be rephrased as follows.
Starting from particle fugacity $z$ and temperature $T$,
we introduce a new couple $(\gamma,T)$ of
independent thermodynamic parameters defined
through the relation $z$ equal a constant times $\gamma\exp ( \beta E_H)$.
Then, we set $T \to 0$ at fixed $\gamma$. The behaviours of graphs in representation
\eqref{SCEdensity} result from the
competition between three mechanisms, which can be roughly
described as follows.

\bigskip

\noindent $\bullet$ \textsl{Screening} : Contributions of bonds 
$-\beta \Phi$, $\beta^2 \Phi^2/2$ and $-\beta^3 \Phi^3/6$ 
are controlled by the inverse Debye screening length
$\kappa=(8\pi\beta e^2 \rho^* \gamma)^{1/2}$ for
ionized protons and ionized electrons with density $\rho^* \gamma$.
They behave as positive or negative powers of $\kappa$, which itself
decays exponentially fast as $\exp ( \beta E_H/2)$.

\bigskip

\noindent $\bullet$ \textsl{Recombination} :
For a particle-cluster made with $N_p$ protons and $N_e$ electrons,
the behaviour of its statistical weight gives raise
to a cluster partition function $Z(N_p,N_e)$ in the vacuum. Each
$Z(N_p,N_e)$ is a truncated trace of Gibbs operator $\exp(-\beta H_{N_p,N_e})$
involving only bare Coulomb Hamiltonians, which
converges thanks to a truncation inherited from screening by
ionized charges~\cite{Alastuey03}. In $Z(N_p,N_e)$,
contributions of all possible recombined entities
made with $M_p \leq N_p$ protons and $M_e \leq N_e$ electrons,
are mixed together. Remarkably, the contribution of
a given chemical species made with $N_p$ protons and $N_e$ electrons,
naturally emerges through Boltzmann factor
$\exp(-\beta E_{N_p,N_e}^{(0)})$, where $E_{N_p,N_e}^{(0)}$
is the groundstate energy of Hamiltonian $H_{N_p,N_e}$. That factor
increases exponentially fast since $E_{N_p,N_e}^{(0)} < 0$.

\bigskip

\noindent $\bullet$ \textsl{Entropy} : In a given cluster,
the presence of $N=N_p+N_e$ particles generates activity powers
$z^N$, which decay exponentially fast as $\exp (N\beta E_H)$.

\bigskip

The behaviour of a graph in the scaling limit is obtained, roughly speaking,
by taking the product of the exponential factors
generated by each of the above mechanisms. Then,
every graph is found to decay exponentially fast. The leading
contributions arise from the first two graphs in
representation \eqref{SCEdensity}. They do reduce to the ideal terms $\rho_p^{\rm id}$ and $\rho_H^{\rm id}$ predicted by Saha theory.
Further corrections decay exponentially faster than $\rho^{\ast}$
in agreement with rigorous bounds \cite{Macris90}.
Dividing all terms by $\rho^{\ast}$, the SLT expansion of 
$\rho /\rho^{\ast}$ reads~\cite{Alastuey08}
\begin{equation}
\label{SLTdensity}
\rho /\rho^{\ast} = \gamma + {\gamma^2 \over 2} +
\sum_{k=1}^{\infty}  \gamma^{n_k} h_k(\beta)~,
\end{equation}
where the first two terms account for contributions from 
ionized particles and hydrogen atoms in their ground state. 
The remaining terms involve functions $h_k(\beta)$ that decay
exponentially fast when $T \to 0$,
while $\gamma^{n_k}$ is an integer or half-integer power of $\gamma$ (which may be multiplied by integer powers of $\ln \gamma$ when $k\geq 6$).
Expansion
\eqref{SLTdensity} is ordered with respect to increasing decay rates
of the $h_k$'s. The corresponding hierarchy
follows from subtle inequalities
between the groundstate energies of all Coulomb
Hamiltonians $H_{N_p,N_e}$. For instance, the molecular contribution,
which determines the leading low-temperature behaviour of
$h_2$, indeed decays exponentially fast thanks to $E_{H_2}> 3E_H$ :
this ensures that molecules $H_2$ are very scarce in the Saha regime
compared to atoms $H$, despite they are more stable
energetically, \textsl{i.e.} $E_{H_2}< 2E_H$.

\section{Equation of state beyond Saha theory} \label{Sct4}

\subsection{Scaled low-temperature expansion of pressure}	

Representation \eqref{SLTdensity} expresses the density in terms of variables $T$ and $\gamma$, or equivalently $T$ and $\mu$ since there is a one-to-one correspondence between those sets of variables. 
As the natural thermodynamical parameters are
the temperature and the density, it is quite useful to invert the
SLT expansion \eqref{SLTdensity} to determine
$\gamma(\rho,T)$, in order to compute all thermodynamical
quantities as functions of $T$ and $\rho$. In the present
scaling limit, this can be done in a perturbative
way around Saha expression
\begin{equation}
\label{gamma_Saha}
\gamma_{Saha}(\rho,T) = \sqrt{1+2\rho/\rho^*} - 1,
\end{equation}
easily obtained by keeping only the first two terms in \eqref{SLTdensity}.
Each correction to that form reduces to a product
of an algebraic function of $\rho/\rho^*$ times a temperature-dependent
function which decays exponentially fast.

\bigskip

The standard thermodynamical relation, which expresses density
in terms of the partial derivative of pressure with respect to $z$ at
fixed $T$, can be rewritten here as
\begin{equation}
\label{pressuregamma}
\rho=  {\gamma \over 2} {\partial \beta P \over \partial \gamma} (\beta,\gamma).
\end{equation}
After inserting SLT expansion \eqref{SLTdensity} of
$\rho$ into \eqref{pressuregamma}, a straightforward integration
with respect to $\gamma$ provides the SLT expansion of
$P$ in terms of $\gamma$ and $T$. The corresponding expansion
\eqref{SLTpressure} of $\beta P/\rho^*$ in terms of $\rho/\rho^*$ and $T$,
then follows by using the inversion relation $\gamma(\rho,T)$
determined above. The physical content of first five corrections
in \eqref{SLTpressure}, as well as the expressions and values of the corresponding
decay rates are summarized in the following table

\bigskip

\begin{tabular}{|c|l|r|}
\hline
Correction ($k$)	&	Physical content	& $\delta_k$ (in eV) \\ \hline
1 & plasma polarization around ionized charges & $|E_H|/2 \simeq 6.8$ \\
2 & formation of molecules, atom-atom interactions & $|3E_H-E_{H_2}| \simeq 9.1$ \\
3 & atomic excitations, charge-charge interactions & $3|E_H|/4 \simeq 10.2$ \\
4 & formation of ions, atom-charge interactions & $|2E_H-E_{H_2^+}| \simeq 11.0$ \\
5 & fluctuations of plasma polarization & $|E_H| \simeq 13.6$ \\ \hline
\end{tabular}

\bigskip

\noindent First correction $k=1$
is equivalent to a modification of Saha ionization
equilibrium~\cite{KreKraLam} derived within
Green functions techniques (see also~Ref.~\cite{Kraeft}), where
rate $\delta_1=|E_H|/2 $ arises from
the behaviour $\kappa \sim \exp (-\beta |E_H|/2) $ in the scaling limit.
All further corrections are entirely new, as well as
the structure of SLT expansion \eqref{SLTpressure}. Beyond their leading
behaviours $\alpha_k(\beta) \sim \exp(-\beta \delta_k)$, functions
$\alpha_k(\beta)$ include further corrections which decay exponentially
faster. For instance, if the leading behaviour of $\alpha_3(\beta)$
is controlled by the first atomic excited state, all the other
contributions of excited states are incorporated into $\alpha_3(\beta)$.
Similarly, $\alpha_2(\beta)$ includes not only the contribution
of the molecular groundstate, but also all contributions of molecular
excited states. As mentioned above about recombination,
the sums of all those contributions are indeed finite.
Also, we stress that such contributions of recombined entities,
like atoms $H$ in $k=3$ or molecules $H_2$ in $k=2$,
are entangled with that of their dissociation products.
Thus, purely atomic or molecular contributions cannot be
unambiguously defined. This does not any trouble here, since
only full contributions embedded in $k=3$ and $k=2$
are relevant for thermodynamics. In other approaches
based on the chemical picture, that ambiguity has
been the source of many controversies since the
introduction of Planck-Larkin formula
(see e.g. Ref.~\cite{Iossilevski}, Ref.~\cite{Norman}
and Ref.~\cite{Starostin}). Eventually,
notice that the contributions to expansion \eqref{SLTpressure}
of more complex entitites, like $H_2^-$, $H_3^+$ or
$H_3$, decay exponentially faster than $\exp (-\beta |E_H|) $
as detailed in Ref.~\cite{Alastuey08}.

\subsection{Low-temperature isotherms}

Let us consider now a small fixed temperature $T$, and
study the behaviour of various corrections to Saha
pressure in \eqref{SLTpressure} when density $\rho$ is varied.
The corresponding low ($\rho \ll \rho^*$) and large
($\rho \gg \rho^*$ ) density behaviours are summarized below :

\bigskip

\begin{tabular}{|c|c|c|}
\hline
$k$ & $\rho \ll \rho^\ast$ & $\rho \gg \rho^\ast$\\ \hline
1 & $\rho^{3/2}$ & $\rho^{3/4}$ \\
2 & $\rho^4$ &  $\rho^2$ \\
3 & $\rho^2$ & $\rho^{1/2}$ \\
4 & $\rho^3$ & $\rho^{3/2}$ \\
5 &  $\rho^2$ & $\rho^{1/2}$ \\
\hline
\end{tabular}

\bigskip

At low densities $\rho \ll \rho^*$, the leading correction of order
$\rho^{3/2}$ is given by plasma polarization (term $k=1$),
while at large densities $\rho \gg \rho^*$ the leading correction
of order $\rho^2$ arises from both molecules $H_2$
and atom-atom interactions (term $k=2$). We have checked that the familiar virial
expansion at low densities is indeed recovered up
to order $\rho^2$ included, since all terms $k \geq 6$
provide powers higher than $\rho^2$. We stress that at too large densities,
expansion \eqref{SLTpressure} breaks down because
various corrections, in particular those due to molecular recombination,
prevail over Saha pressure which grows only as $\rho$. In fact,
a semi-empirical criterion based on that observation allows us
to infer a validity domain for the SLT expansion, as described below.

\section{Numerical applications and comparisons to Monte Carlo data}	

The truncated EOS obtained by keeping the first
five terms in SLT expansion \eqref{SLTpressure} can be computed numerically.
The $b_k$'s are easily computed since they reduce
to simple algebraic functions of ratio $\rho/\rho^{\ast}$.
Numerical values for temperature-dependent functions
$\alpha_1$, $\alpha_3$ and $\alpha_5$ are also readily inferred
from explicit analytic expressions. No similar expressions for
$\alpha_2$ and $\alpha_4$ are available, since
analytical results on the three- and four-body
quantum problem are very scarce. Then, we use simple modelizations
of those functions which account for their exact low-temperature
forms on one hand, and incorporate familiar phenomenological
descriptions of ions $H^-$ and $H_2^+$ and of molecule $H_2$
on the other hand~\cite{AlaBalCor09}.

\bigskip

Various isotherms corresponding to increasing
temperatures have been considered. For temperatures below $2000 K$,
the Saha regime defines extremely diluted conditions which do not
make physical sense : this explains why Earth or Brown Dwarfs
atmospheres only involves molecules $H_2$. Temperature
can be increased up to $T=30000K$, which is still small
compared to the characteristic temperature
scale $|E_H|/k_B \simeq 150 000K$. At a given temperature,
calculations within the truncated EOS are not reliable above some
density $\rho_c$, for which corrections to Saha pressure
become too large, in agreement with previous estimations
for $\rho \gg \rho^*$. That breakdown is due to molecular recombination
below $T \simeq 16000$~K, and to atom-atom interactions for
higher temperatures.

At relatively low temperatures, i.e. below $T=10000K$,
PIMC calculations \cite{Militzer01} have been performed at rather high densities
$\rho >> \rho_c$ for which atoms $H$ are mainly recombined into
molecules $H_2$. In the Saha regime, statistics in PIMC results are
poor because the corresponding densities are too diluted. In fact,
under such conditions, our analytical results might serve as testbenchs
for simulation methods. For higher temperatures, i.e.
above $T=10000K$, there exists a density range (see PIMC crosses
displayed in Fig.~\textbf{validitydomain}), where comparisons between
our results and PIMC data \cite{Militzer01} are instructive. A good
agreement is observed in some density range, which can be
inferred from a semi-empirical criterion : corrections  cannot exceed
a few per cent of Saha pressure. This defines, at a heuristic level,
the validity domain of SLT expansion shown in Fig.~\textbf{validitydomain}.
In that whole domain, weak-coupling and weak-degeneracy conditions
are fullfilled. The tongue structure of the domain between $T \simeq 10000$~K
and $T \simeq 25000$~K results from the
increase of the strength of interactions between ionized charges and atoms.
Notice that the whole domain is restricted to rather low densities in
general, so high-density phenomena, like the celebrated plasma phase
transition, remain beyond the scope of our approach.

\bigskip

Eventually, we emphasize that our numerical calculations will be
detailed in a forthcoming paper~\cite{AlaBalCor09}, where
simple representations of functions $\alpha_k(\beta)$
($k=1,...,5$) will be given, while the
corresponding functions $b_k(\rho/\rho^{\ast})$
can already be found in Ref.\cite{Alastuey08}. Our results will be
compared to PIMC data, and also to phenomenological calculations.
Applications to the Sun adiabat should be considered later.

\end{document}